\documentclass[manuscript]{acmart}

\makeatletter
  \@twosidetrue  
  \@twosidefalse  
  \@mparswitchfalse
\makeatother

\AtBeginDocument{%
  \providecommand\BibTeX{{%
    Bib\TeX}}}

\setcopyright{acmlicensed}
\copyrightyear{2025}
\acmYear{2025}
\acmDOI{10.1145/3798164}
\acmISBN{978-1-4503-XXXX-X/2018/06}

\usepackage{hyperref}
\usepackage{graphicx}
\usepackage{textcomp}
\usepackage{enumitem}
\usepackage{mathtools}
\usepackage{multirow}
\usepackage[most]{tcolorbox}
\usepackage{algorithm}
\usepackage{algorithmic}
\usepackage{url}
\definecolor{lavender}{rgb}{0.8, 0.7, 1}
\usepackage[table]{xcolor}
\usepackage{fancyvrb}
\tcbuselibrary{listings, breakable}
\usepackage{fancyvrb}
\usepackage{pifont}

\newcommand{\nff}[1]{\textcolor{black}{#1}}

\newtcbox{\myverbatimbox}[1][]{nobeforeafter, colback=gray!5, colframe=black!50,
  fontupper=\ttfamily, boxrule=0.5pt, arc=2pt, outer arc=2pt,
  top=3pt, bottom=3pt, left=5pt, right=5pt, enhanced,
  sharp corners=south, breakable, listing only, listing options={
    basicstyle=\ttfamily\small,
    breaklines=true,
    commandchars=\\\{\}
  }, #1}

\newcommand{\answer}[1]{
\vspace{2mm}\noindent\fbox{%
    \parbox{.97\columnwidth}{%
        {#1}
    }%
}\vspace{2mm}}

\newtcolorbox{modelbox}[1][]{%
    colback=gray!10,            
    colframe=black!100,           
    boxrule=0.1mm,              
    fonttitle=\bfseries,         
    coltitle=black,             
    enhanced,                    
    before skip=10pt,           
    after skip=10pt,           
    parskip=false,  
    attach boxed title to top left={xshift=3mm, yshift=-2mm, yshifttext=-1mm}, 
    boxed title style={%
        colback=green!60!black, 
        colframe=black!100, 
        boxrule=0.1mm,         
        arc=2pt,               
        width=3cm,             
    },
    title={\textcolor{white}{\textbf{\faKeyboardO~Input Prompt}}}, 
    #1
}

\newtcolorbox{modelboxoutput}[1][]{%
    colback=gray!10,            
    colframe=black!100,           
    boxrule=0.1mm,               
    fonttitle=\bfseries,         
    coltitle=black,              
    enhanced,                    
    before skip=3pt,            
    after skip=3pt,             
    attach boxed title to top left={xshift=3mm, yshift=-2mm, yshifttext=-1mm}, 
    boxed title style={%
        colback=orange!90!white,  
        colframe=black!100,
        boxrule=0.1mm,             
        arc=2pt,                 
        width=3cm,              
    },
    title={\textcolor{white}{\textbf{\faPrint~Model Response}}}, 
    #1
}

\def\BibTeX{{\rm B\kern-.05em{\sc i\kern-.025em b}\kern-.08em
    T\kern-.1667em\lower.7ex\hbox{E}\kern-.125emX}}

\newcommand{\ourapproach}{\mbox{RTM}} 

\begin{document}

\title{Requirements Coverage-Guided Minimization for Natural Language Test Cases}

\author{Rongqi Pan}
\email{rpan@uottawa.ca}
\orcid{0000-0002-9096-6241}
\affiliation{%
  \institution{University of Ottawa}
  \city{Ottawa}
  \state{Ontario}
  \country{Canada}
}

\author{Feifei Niu}
\email{feifeiniu96@gmail.com}
\orcid{0000-0002-4123-4554}
\affiliation{%
  \institution{University of Ottawa}
  \city{Ottawa}
  \state{Ontario}
  \country{Canada}
}

\author{Lionel C. Briand}
\affiliation{%
  \institution{University of Ottawa}
  \state{Ontario}
  \city{Ottawa}
  \country{Canada}}
  \affiliation{%
  \institution{Research Ireland Lero Centre, University of Limerick}
  \city{Limerick}
  \country{Ireland}}
\email{lbriand@uottawa.ca}
\orcid{0000-0002-1393-1010}

\author{Hanyang Hu}
\orcid{0009-0007-3878-2430}
\affiliation{%
  \institution{Wind River Systems}
  \country{Canada}
}
\email{phenom.hu@gmail.com}

\renewcommand{\shortauthors}{Pan et al.}

\begin{abstract}
As software systems evolve, test suites tend to grow in size and often contain redundant test cases. Such redundancy increases testing effort, time, and cost. Test suite minimization (TSM) aims to eliminate such redundancy while preserving key properties, such as requirement coverage and fault-detection capability. 
In this paper, we propose \ourapproach~(Requirement coverage-guided Test suite Minimization), a novel TSM approach designed for requirement-based testing (validation), which can effectively reduce test suite redundancy while ensuring full requirement coverage and a high fault detection rate (\textit{FDR}) under a fixed minimization budget. Based on common practice in critical systems where functional safety is important, we assume that test cases are specified in natural language and traced to requirements before implementation. 
\textcolor{black}{\ourapproach~utilizes text embedding technique to convert test cases into vector representations, on which a distance function is employed to compute similarity values between test case pairs. Guided by these similarity values, a Genetic Algorithm (GA) whose population is initialized using a coverage-preserving strategy is then employed to search for an optimal subset of diverse test cases that matches the budget. We investigate three preprocessing methods for test cases, seven different text embedding techniques, three distance functions, and three  initialization strategies for the GA.}
We evaluate \ourapproach~on an industrial automotive system dataset comprising $736$ system test cases covering $54$ requirements. Experimental results show that, \textcolor{black}{while being scalable in terms of runtime, \ourapproach~outperforms all the baseline techniques in terms of \textit{FDR} on most minimization budgets while maintaining full requirement coverage.} Furthermore, we investigate the impact of test suite redundancy levels on the effectiveness of TSM, providing new insights into optimizing requirement-based test suites under practical constraints.

\end{abstract}

\setcopyright{cc}
\setcctype{by-nc-nd}
\acmJournal{TOSEM}
\acmYear{2026} \acmVolume{1} \acmNumber{1} \acmArticle{}
\acmMonth{1} \acmDOI{10.1145/3798164}




\received{8 May 2025}
\received[revised]{30 Novermber 2025}
\received[accepted]{16 February 2026}

\maketitle

\section{Introduction}
Software testing is a critical activity throughout the software development life cycle, playing a key role in early defect detection and in ensuring requirements compliance~\cite {myers2011art,everett2007software}. In particular, requirements-driven testing is conducted to validate systems before deployment~\cite{dustin2009implementing}. This is especially critical for systems that must demonstrate functional safety, such as automotive systems. In such contexts, it is also common to specify test cases in natural language (following templates) to facilitate their analysis and traceability to requirements, before manually writing the corresponding test code. 

However, due to the increasing complexity of systems and their continuous change by large teams, test suites are hard to maintain, tend to grow, and are often redundant (i.e., some of the test cases are likely to detect the same faults), thus leading to waste of time and  resources~\cite{rothermel2002empirical}. Further, in practice, testers have limited resources and time to test new system updates. To address this, given a test budget, test suite minimization (TSM) is proposed to prune test cases that are most likely to be redundant, while satisfying coverage criteria and maximizing fault detection~\cite{harrold1993methodology, von1999test}, thus significantly reducing testing cost and time while limiting the impact on system quality or compliance.

While several approaches have been proposed for TSM~\cite{noemmer2019evaluation, miranda2017scope, coviello2018clustering, liu2011user, viggiato2022identifying, cruciani2019scalable, zhang2019uncertainty, hemmati2013achieving, pan2023atm, DBLP:journals/tse/PanGB24, marchetto2017combining, xia2021test, anwar2019hybrid, yoo2010using}, most existing techniques rely on code and structural coverage criteria (e.g., statement or branch coverage). These techniques do not address the unique challenges of minimizing test suites driven by system requirements, with test cases often specified in natural language. Moreover, to the best of our knowledge, no existing solution can simultaneously maintain full requirement coverage while adhering to a fixed minimization budget, a common situation in industrial practice due to limited time and resources. 

To bridge this gap, we propose \ourapproach~(Requirements coverage-guided Test suite Minimization), a novel approach specifically designed for requirement-based testing. \ourapproach~aims to reduce test suite redundancy (in terms of fault detection) while ensuring full requirement coverage, making it particularly well-suited to the practical needs of our industrial partners and to critical systems that must demonstrate functional safety in general. \ourapproach~takes requirement-based test cases--composed of natural language test steps--as input and computes pairwise similarity between them. Specifically, we investigate three different preprocessing methods to normalize the test steps, followed by seven distinct text embedding techniques--both sentence-level and word-level--to transform the textual data into vector representations: Term Frequency–Inverse Document Frequency (TF-IDF)~\cite{salton1988term}, Universal Sentence Encoder (USE)~\cite{DBLP:journals/corr/abs-1803-11175}, LongT5~\cite{DBLP:conf/naacl/GuoAUONSY22}, Amazon Titan Text Embedding V2~\cite{aws2024titantv2}, Word2Vec~\cite{DBLP:journals/corr/abs-1301-3781}, GloVe~\cite{DBLP:conf/emnlp/PenningtonSM14}, FastText~\cite{DBLP:journals/tacl/BojanowskiGJM17}. We then measure the similarity between test cases using three different distance metrics: cosine similarity and Euclidean distance for sentence-level embeddings, and Word Mover’s Distance (WMD)~\cite{DBLP:conf/icml/KusnerSKW15} for word-level embeddings.
Based on the computed similarities, we employ a Genetic Algorithm (GA)~\cite{Luke2013Metaheuristics} to search for the optimal test suite that minimizes similarity, adheres to a fixed budget, and ensures full requirement coverage. To help the GA efficiently find valid subsets, we devise three initialization strategies that generate an initial population of subsets satisfying both coverage and budget constraints.

\ourapproach~was evaluated on an industrial automotive dataset consisting of $736$ system test cases across seven test runs, covering $54$ requirements. 
The results demonstrate that \ourapproach~outperforms the baselines in terms of fault detection rate (\textit{FDR}) under various minimization budgets.
Moreover, we explore the impact of test suite redundancy on \textit{FDR}. Our findings show that: \ding{172}
\ourapproach~consistently outperforms all baseline techniques in terms of \textit{FDR} across different test suite redundancy levels;
\ding{173} The level of redundancy has a strong effect on the achieved \textit{FDR}, for all TSM techniques;
\ding{174} Especially at lower minimization budgets and for highly redundant test suites, there is still significant room for improvement between the theoretical maximum \textit{FDR} and that of \ourapproach. \textcolor{black}{\ding{175}  The runtime of \ourapproach~scales approximately linearly with test suite size, with small (i.e., of no practical consequence) absolute differences compared to the baselines. Given its consistently higher \textit{FDR} performance across various test suites with different redundancy levels, \ourapproach~is therefore a better choice in many practical contexts.}

In summary, our contributions include:
\begin{itemize}
    \item We propose \ourapproach, a TSM approach tailored for requirement-based testing, where test cases are specified in natural language. \ourapproach~ensures full requirement coverage while operating under a fixed minimization budget, addressing a practical constraint often overlooked in prior work.

    \item We validate \ourapproach~on automotive system requirements and test cases, and compare it with several baseline techniques for test suite minimization across varying minimization budgets, demonstrating its superior fault detection capability.
 
    \item We further investigate the impact of test suite redundancy levels on TSM effectiveness, offering new insights into the relationship between test suite redundancy and TSM fault-detection performance.

    \item To foster reproducibility and facilitate further research, we have open-sourced our replication package, including source code, \textcolor{black}{input data, and results for all research questions that allow the reproducibility of the experiments}, which is publicly available at \url{https://github.com/rongqipan/RTM}. Industrial requirements and test cases are confidential. 
\end{itemize}

The rest of the paper is organized as follows: Section~\ref{sec:problemdefinition} defines the TSM problem and outlines the context of this study. Section~\ref{sec:approach} describes our approach: \ourapproach. The study design is detailed in Section~\ref{sec:studydesign}, followed by experimental results in Section~\ref{sec:results}. Section~\ref{sec:threats} discusses threats to validity. Section~\ref{sec:relatedwork} reviews related work. Finally, Section~\ref{sec:conclusion} concludes the study with a discussion of future work. 

\section{Problem Definition: Test Suite Minimization in Automotive Requirements Testing} \label{sec:problemdefinition}
In this paper, we aim to minimize test suites derived from automotive system requirements. In this section, we first describe the format and structure of the test cases in our dataset, followed by a formal definition of the TSM problem in our specific context.

\subsection{Test Cases based on Automotive Requirements}
\label{sec:test_case}

Like other critical systems, modern automotive software systems are expected to comply with functional safety standards such as ISO~26262~\cite{siegl2010model}. To ensure compliance, system engineers derive and maintain extensive test suites based on system requirements, which are executed to validate system behavior against expected requirements. Typically, test cases are specified in natural language, following templates, and then manually converted into test code and implemented by experts. 

The test cases in our dataset were collected from our industry partner and designed to test automotive system requirements. As shown in Figure~\ref{fig:tc_example}, these test cases are specified in natural language. Each test case consists of a sequence of diagnostic actions, structured into clearly labeled steps. These steps typically include setting up the preconditions for testing, creating the fault conditions, checking how the automotive system detects the issues through Diagnostic Trouble Codes (DTCs), and confirming that the system properly clears the DTC.
The detailed test actions involve reading and setting system variables, awaiting signal conditions, sending diagnostic requests, and checking system responses.
To this end, the test cases exhibit the following characteristics:

\noindent
\textbf{Structured and Precise}: Test cases explicitly define preconditions, test steps, and expected outcomes.

\noindent
\textbf{Domain-Specific Vocabulary}: Test cases extensively utilize automotive-specific terminology, signals, variables, and abbreviations.

However, as the number of system requirements increases, the corresponding test suites can become excessively large, resulting in dramatically higher testing costs. Moreover, as new requirements are added, test cases are often written without checking the existing ones. This can lead to functional overlap between test cases, thus increasing redundancy in the test suite.

\begin{figure}[htbp]
\centering
{\fontsize{8pt}{8pt}\selectfont
\begin{tcolorbox}[colframe=black!80!white, colback=gray!5, boxrule=0.5pt, sharp corners, width=0.5\linewidth]
\begin{Verbatim}[commandchars=\\\{\}]
\textbf{STEP 1 Set Global Preconditions}
\textbf{Read} variable \textit{Variable_A}
\textbf{STEP 2 Set Valid Preconditions}
\textbf{Set} System variable \textit{Variable_1} = 1
\textbf{Await} Value Match Signal \textit{SIGNAL_A} = 1
\textbf{STEP 3 Create Fault Condition}
\textbf{Set} System variable \textit{Variable_2} = 1
\textbf{STEP 4 Verify DTC Maturation Time}
\textbf{Check} maturation time (Expected: \textit{X} ms)
\textbf{STEP 5 Check the DTC is Active}
\textbf{Read} variable \textit{Variable_A}
\textbf{Send} request \textit{PATH\_TO\_REQUEST\_A}
\textbf{Check} expected diagnostic response
\textbf{Set} System variable \textit{Variable_1} = 0
\textbf{Await} Value Match Signal \textit{Variable_1} = 0
\textbf{STEP 6 Remove DTC Condition}
\textbf{Set} System variable \textit{Variable_2} = 0
\textbf{STEP 7 Verify DTC Dematuration Time}
\textbf{Read} variable \textit{Variable\_A}
\textbf{Check} dematuration time (Expected: \textit{X} ms)
\textbf{Send} request \textit{PATH\_TO\_REQUEST\_A}
\textbf{Check} expected diagnostic response
\end{Verbatim}
\end{tcolorbox}
}
\caption{A sanitized example of the test case in our dataset}
\label{fig:tc_example}
\end{figure}

\subsection{Test Suite Minimization}\label{sec:definition}

Test suite minimization (TSM) aims to reduce the test suite size while ensuring key testing objectives, such as requirement coverage, fault-detection capability, and test-case diversity~\cite{harrold1993methodology, von1999test}. Given that test suite diversity is reported to be positively correlated with fault detection capability~\cite{hemmati2013achieving}, minimizing test case similarity can enhance the fault detection capability of the resulting test suite.
When appropriately applied, TSM can significantly reduce testing costs and time while minimizing its impact on system quality or compliance.

Let $\mathcal{R} = \{r_1, r_2, \dots, r_n\}$ be the set of system requirements for an automotive system, and let $\mathcal{T} = \{t_1, t_2, \dots, t_m\}$ be the corresponding test suite. Each test case $t_i \in \mathcal{T}$ covers a subset of requirements denoted by $Cover(t_i) \subseteq \mathcal{R}$. Note that in our dataset, each test case only covers one requirement, whereas many test cases can cover each requirement. This is because each test case is specifically designed to validate a scenario in which the requirement is expected to hold. \textcolor{black}{Therefore, in our context, guaranteeing full requirement coverage requires at least as many test cases as requirements in the test suites.} Moreover, covering all requirements is a must in automotive testing, to comply with functional safety standards~\cite{siegl2010model}, which are adhered to by our industry partner. Let $Sim(t_i, t_j): \mathcal{T} \times \mathcal{T} \rightarrow [0,1]$ denote the similarity between test cases $t_i$ and $t_j$.

Specifically, TSM in our context can be defined as the following optimization problem:

\noindent
\textbf{Given:} test suite $\mathcal{T}$, requirement set $\mathcal{R}$ with $r$ requirements, budget ${\color{black}M} \in (0,1]$,
coverage function $Cover(t) \subseteq \mathcal{R}$ for each $t \in \mathcal{T}$, and similarity function $Sim(t_i, t_j)$.

\noindent
\textbf{Find:} a subset $\mathcal{T}' \subseteq \mathcal{T}$ such that:

\begin{alignat}{2}
& \textcolor{black}{|\mathcal{T}'|} && \textcolor{black}{= \max\!\bigl\{r,\ \lfloor M\,|\mathcal{T}|\rfloor \bigr\}} \label{eq:size}\\
& \bigcup_{t \in \mathcal{T}'} Cover(t) &&= \mathcal{R}, \label{eq:coverage} \\
& \sum_{\substack{t_i, t_j \in \mathcal{T}'\\i < j}} Sim(t_i, t_j) &&\text{ is minimized.} \label{eq:similarity}
\end{alignat}

This problem extends the classical \emph{set cover problem}, which is known to be NP-hard, by introducing two additional constraints: a strict minimization budget and a diversity objective. In particular, the selected test subset must maintain full requirement coverage within a fixed budget while minimizing overall test-case similarity. As a result, solving this problem exactly is computationally intractable for large-scale industrial systems. Therefore, practical heuristics or approximation strategies are typically employed in automotive domains to achieve efficient and effective test suite minimization, especially when additional factors, such as fault detection capability, execution cost, or coverage redundancy, are also considered.

\section{Approach}\label{sec:approach}

This section introduces \ourapproach, a TSM approach that satisfies the three criteria outlined in Section~\ref{sec:definition}. It targets test cases derived from system requirements for validation. In short, it leverages \textcolor{black}{a text embedding technique combined with a distance function to compute test case similarity, then employs evolutionary search to minimize the test suite. The primary objective of \ourapproach~is to find the optimal $\mathcal{T}'$ that maximizes fault detection capability by minimizing the similarity values between the test cases while reducing the testing cost, measured in terms of test suite size, as well as ensuring $100\%$ requirement coverage.}

\begin{figure*}[htbp]
\centering
\includegraphics[width=0.9\textwidth]{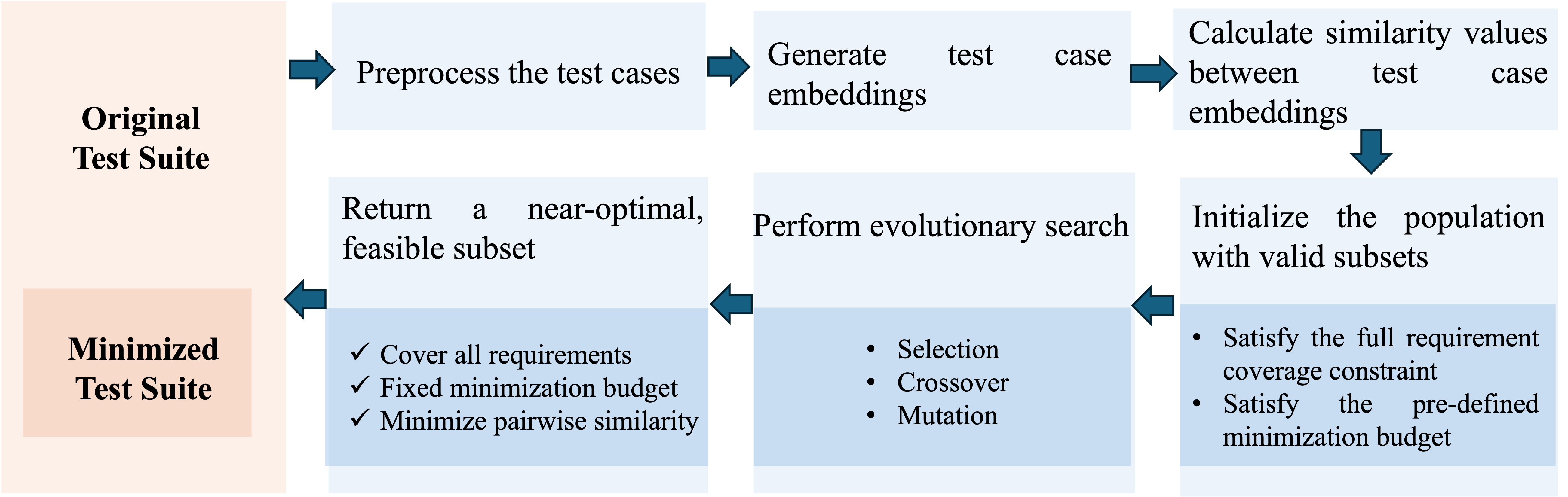}
\Description{}
\caption{Overall framework of \ourapproach.}
\label{fig:approach}
\end{figure*}

\textcolor{black}{Figure~\ref{fig:approach} outlines the main steps of \ourapproach. We first describe how we preprocess test cases (Section~\ref{sec:preprocessing}) and then convert them into vector representations (Section~\ref{sec:representation}). Then, we describe the distance function employed to calculate similarity values between test case embeddings (Section~\ref{sec:similarity}). Finally, we describe the search algorithm (GA) we employed to search for the near-optimal minimized test suite containing diverse test cases while satisfying all the constraints (Section~\ref{sec:search_algorithm}).}

\subsection{Test Case Preprocessing}
\label{sec:preprocessing}
Given that the test cases are expressed in natural language, different preprocessing strategies may affect the outcomes of the text embedding techniques. \textcolor{black}{We investigated three different levels of preprocessing:}

\begin{itemize}
    \item \textbf{Preprocessing Method 1 ($PM1$)} converts all characters to lowercase, removes extra white spaces between words, and eliminates line breaks (e.g., \verb|\r\n|).
    \item \textbf{Preprocessing Method 2 ($PM2$)} extends \textit{$PM1$} by further stripping punctuation and performing tokenization and lemmatization.
    \item \textbf{Preprocessing Method 3 ($PM3$)} serves as a control setting, where no preprocessing is applied, and the raw test case content is used as-is.
\end{itemize}

\subsection{Test Case Representation} 
\label{sec:representation}
After the preprocessing step, we transform the \textcolor{black}{preprocessed test cases} into vector representations using a text embedding technique. We \textcolor{black}{investigate} both word-level and sentence-level embedding techniques to capture semantic information from test cases. 

For word-level embeddings, we use Word2Vec, GloVe, and FastText,  \nff{which maps each word in a test case to a $300$ dimensional numerical vector. Consequently, a test case with $k$ words is represented as a $k\times300$ matrix. Note that we do not aggregate word vectors into a single sentence embedding. We retain the word-level embeddings, based on which the Word Mover’s Distance (WMD) is computed, as described in the following section.}

For sentence-level embeddings, we employ TF-IDF, USE, LongT5, and Amazon Titan Embedding V2, \nff{each of which encodes the entire test case into a single fixed-length vector that captures the overall semantics of the entire test case content. The dimensionality of these vectors varies across embedding techniques. For example, TF-IDF produces sparse vectors whose dimensionality equals the vocabulary size, USE yields a $512$-dimensional vector, LongT5 produces a $768$-dimensional embedding, and Amazon Titan Embedding V2 returns a $1024$-dimensional vector. These differences do not affect our approach, as similarity, as described in the following section, is computed independently for each embedding technique without requiring dimensional alignment.}

\subsection{Similarity Measure} \label{sec:similarity}
To assess the similarity between test case embeddings, we \textcolor{black} {investigate} different distance metrics depending on the embedding type.

For word-level embeddings (i.e., Word2Vec, GloVe, and FastText), we use WMD~\cite{DBLP:conf/icml/KusnerSKW15}, which \nff{formulates the similarity between two test cases as an optimal transport problem. Specifically, WMD computes the minimum cumulative cost required to move the probability mass of all the words in one test case to those in another in the embedding space. The WMD is defined as:}

\nff{
\begin{equation}
\mathrm{WMD}(\mathbf{t}_1, \mathbf{t}_2) = 
\min_{T \ge 0} \sum_{i,j=1}^{n} T_{ij} \, d(i,j)
\label{eq:wmd}
\end{equation}
}

\noindent \textcolor{black}{where $n$ is the global vocabulary size, $T_{ij}$ denotes how much of the probability mass of word $i$ in test case $t_1$ moves to word $j$ in test case $t_2$, and $d(i,j)$ represents the Euclidean distance between word embeddings $i$ and $j$. If a word $i$ appears $f_i$ times in test case $t_1$, we denote the probability mass of word $i$ as $p_{1i} = \frac{f_i}{\sum_{j=1}^{n} f_j}$. The transport plan $T$ (a matrix) is subject to the constraints $\sum_j T_{ij} = p_{1i}$ and $\sum_i T_{ij} = p_{2j}$, $\forall i,j\in\{1,\dots,n\}$, which ensure test case $t_1$ is entirely transformed into test case $t_2$. } 

\nff{This optimization problem is equivalent to the Earth Mover’s Distance~\cite{monge1781memoire, wolsey1999integer, rubner1998metric} and can be solved via linear programming, 
with a computational complexity of approximately $O(n^3 \log n)$ for $n$ unique words. 
Although the computation can be expensive when there are many words,} 
WMD is well-suited to word-level embeddings because pairwise distances between the word embeddings approximate semantic dissimilarity, allowing the transport cost to reflect meaning differences between test cases~\cite{DBLP:conf/icml/KusnerSKW15}.

\textcolor{black}{In order to bound the WMD between $0$ and $1$, and convert it from a distance function to a similarity measure, we normalize it as follows:}

\begingroup
\color{black}
\begin{equation}
\label{eq:wmd-norm}
\operatorname{WMD}_{\text{norm}}(\mathbf{t}_1,\mathbf{t}_2)
= \frac{1}{1+\operatorname{WMD}(\mathbf{t}_1,\mathbf{t}_2)} \in (0,1].
\end{equation}
\endgroup

For sentence-level embeddings (i.e., TF-IDF, USE, LongT5, and Amazon Titan Embedding V2), we \textcolor{black}{investigate} both cosine similarity and Euclidean distance, which have been widely used in machine learning, information retrieval, and NLP tasks to measure similarity or dissimilarity between vectors~\cite{gomaa2013survey, huang2008similarity, li2013distance, salton1988term}. However, they operate differently and are suited to different use cases: cosine similarity measures the angle between two vectors, ignoring magnitude, whereas Euclidean distance measures the straight-line distance between two vectors. By applying different types of similarity metrics, we aim to explore how different vector representations and distance functions affect the ability of \ourapproach~to detect redundant test cases.

\textcolor{black}{We normalize the cosine similarity and Euclidean distance as follows: }

\begingroup
\color{black}
\begin{equation}
\label{eq:cos-ang}
\operatorname{Cosine~Similarity}_{\text{norm}}(\mathbf{t}_1,\mathbf{t}_2)
= 1 -  \frac{\arccos{(cosine~similarity(\mathbf{t}_1, \mathbf{t}_2))}}{\pi}
= 1 -  \frac{\arccos{\frac{\mathbf{T_1} \cdot \mathbf{T_2}}{\|\mathbf{T_1}\|\|\mathbf{T_2}\|}}}{\pi} \in [0,1].
\end{equation}
\endgroup
\textcolor{black}{where $T_1$ and $T_2$ are the embeddings of test case $t_1$ and $t_2$, respectively.}

\begingroup
\color{black}
\begin{equation}
\label{eq:l2-sim}
\operatorname{Euclidean~Distance}_{\text{norm}}(\mathbf{t}_1,\mathbf{t}_2)
= \frac{1}{1+Euclidean~Distance (\mathbf{t}_1, \mathbf{t}_2)}
= \frac{1}{1+\sqrt{\sum_{i=1}^n (t_{1i}-t_{2i})^2}} \in (0,1].
\end{equation}
\endgroup
\textcolor{black}{where $t_{1i}$ and $t_{2i}$ denote the $i^{th}$ element in the embeddings of test case $t_1$ and $t_2$, respectively. $n$ is the embedding dimension (i.e., total number of elements). }

This step constructs a similarity matrix by computing the pairwise similarity between test cases. It takes the vector representations of test case pairs as input and calculates the similarity scores \textcolor{black}{using one of the similarity measures} to quantify their similarity.

\subsection{Evolutionary Search for Test Suite Minimization}
\label{sec:search_algorithm}
\textcolor{black}{As discussed in Section~\ref {sec:definition}, the TSM in our context is an NP-hard problem. Therefore, to efficiently find a near-optimal, feasible solution, we apply a GA--the most widely used evolutionary search algorithm in software testing and the approach adopted by the state-of-the-art (SOTA) TSM work (i.e., ATM~\cite{pan2023atm} and LTM~\cite{DBLP:journals/tse/PanGB24})--which has demonstrated high effectiveness and efficiency for TSM. To tailor GA to our problem, we define the individuals (minimized test suites), the initialization strategy, selection, crossover, and mutation operators, and the fitness function that guides the search. The details are given below.}
\subsubsection{Problem Formalization}
\label{sec:problem_formalization}
Given the test suite before minimization $\mathcal{T}$, requirements set $\mathcal{R}$ with $r$ requirements, budget $M \in (0,1]$, the minimization problem is defined as a fixed-size subset selection with $100\%$ requirement coverage as the constraint. Each individual (i.e., subset $\mathcal{T}'$) is represented as a binary vector:

\begin{equation} \label{eq:solution_vector}
    \mathbf{x} = [x_1, x_2, \ldots, x_m] ,\quad 
    x_i = 
    \begin{cases}
        1, & \text{if test case } i \text{ is selected} \\
        0, & \text{otherwise}
    \end{cases}
\end{equation}
where m denotes the total number of test cases in the test suite before minimization (i.e., $\mathcal{T}$).

Given the minimization budget ${\color{black}M} \in (0, 1]$, the number of test cases in the subset $\mathcal{T}'$ is constrained by:
\begin{equation} \label{eq:budget_constraint}
    \sum_{i=1}^m x_i = \max\!\left(r, \operatorname{round}\!\left(m \cdot {\color{black}M}\right)\right)
\end{equation}
where $\mathrm{round}()$ denotes rounding the number to the nearest integer. Note that the number of test cases in $\mathcal{T}'$ should be higher than or equal to the total number of requirements.

Moreover, the test cases in the subset must cover all the requirements:
    \begin{equation}
        \bigcup_{\{i \mid x_i=1\}} Cover(i)  = \mathcal{R}
    \end{equation}
    where $Cover(i) $ is the requirement covered by test case $i$ and $\mathcal{R}$ is the set of requirements covered by the test suite before minimization (i.e., $\mathcal{T}$).

Given that the objective of the search process is to minimize the similarity values between test case pairs in the minimized test suite, the fitness function is formulated as:

\begingroup
\color{black}
\begin{equation}
\label{eq:fitness}
\mathrm{Fitness}(\mathcal{T}'_n)
= \frac{1}{n}\sum_{i=1}^{n}
\biggl( \max_{\substack{1 \le j \le n \\ j \ne i}}
\operatorname{NormSim}(t_i,t_j) \biggr)^{2},
\quad \text{where } \mathcal{T}'_n=\{t_1,\dots,t_n\}.
\end{equation}
where $\mathcal{T}'_n$ denotes the subset containing $n$ test cases, and $\operatorname{NormSim}(t_i,t_j)$ represents the normalized similarity score for the test case pair $(t_i,t_j)$. Individuals with lower fitness values are better, indicating more diverse test cases with lower similarity.
\endgroup

\subsubsection{Initialization strategies} \label{sec:initialization}

Our minimization problem is defined with two hard constraints: (1) the percentage of selected test cases in the subset should satisfy the minimization budget, and (2) the subset should preserve $100\%$ requirement coverage. To efficiently guide the search for valid subsets, we \textcolor{black}{investigate} three strategies to initialize a set of valid subsets for the GA to use as a base for further search. Note that, in our dataset, as is often the case, the relationship between test cases and requirements is many-to-one. Consequently, one \textcolor{black}{experimental setting of our minimization problem is that the pre-defined number of selected test cases (\textcolor{black}{$|\mathcal{T}'|$}) must not be smaller than the total number of requirements (\textcolor{black}{$|\mathcal{R}|$} ), thus satisfying the full requirement coverage constraint when combined with the initialization strategy. }

\begin{itemize}[leftmargin=3em]
    \item \emph{Strategy 1. Iterative Selection.} 
    \begin{itemize}
    \item This strategy iteratively selects one test case randomly under each requirement, continuing this process until the percentage of selected test cases reaches the minimization budget.
    \end{itemize}
    \item \emph{Strategy 2. Initial Requirement + Random.} 
    \begin{itemize}
        \item Step 1: Randomly select one test case for each requirement.
        \item Step 2: Randomly select the remaining test cases without considering requirements until the percentage of selected test cases reaches the minimization budget.
    \end{itemize}
    
    \item \emph{Strategy 3. Proportional Selection.}
    \begin{itemize}
       \item Step 1: Randomly select a proportion of test cases (approximately) equal to the minimization budget under each requirement.
       \item Step 2: If the percentage of selected test cases across all requirements is less than or greater than the minimization budget, iteratively add or reduce the number of selected test cases per requirement, \textcolor{black}{while ensuring every requirement is still covered by at least one test case}, until the desired minimization budget is reached.
    \end{itemize}
    
\end{itemize}

\subsubsection{\color{black}{Search Process}} \label{sec:evolutionary}
\textcolor{black}{After initializing a population with valid individuals (subsets), GA~\cite{golberg1989genetic} iteratively evolves this population using selection, crossover, and mutation operators towards better individuals with lower fitness values. In each generation, offspring are created by recombining and mutating the genes (i.e., selected test cases) of existing individuals. Each individual is evaluated by computing its fitness, and those with lower fitness are selected for the next generation. The evolutionary process is repeated until convergence, defined as a decrease in the fitness score of less than $0.0025$ across generations. The final output is the best subset identified during the search, which satisfies all constraints and minimizes the redundancy among selected test cases, indicated by the lowest fitness value.}

Specifically, we adopt the following three genetic operators:

\begin{itemize}
    \item \textbf{Selection:} We used a binary tournament selection operator that prioritizes valid subsets—those that cover all requirements—over invalid ones. When both candidate subsets are valid, the one with the lower fitness value is selected.
    
    \item \textbf{Crossover:} We used the customized crossover operator~\cite{pymoo} that ensures the offspring maintains a fixed subset size. It first includes all test cases shared by both parent subsets, then adds test cases from either parent subset until the minimization budget is met.
    
    \item \textbf{Mutation:} We used an inversion mutation operator~\cite{pymoo} that randomly selects a segment of the subset and reverses its order. Since a subset is represented as a binary vector, this approach preserves the subset size.
\end{itemize}

Note that crossover and mutation may produce invalid individuals (i.e., those that violate the coverage constraint). To address this, we experimented with incorporating a repair operator during the search process. The repair operator randomly adds one test case for each uncovered requirement and removes the test cases under requirements that are covered by more than one test case, therefore reestablishing full coverage while maintaining the budget constraint for each solution. However, the results obtained using the repair operator were slightly worse than those without it. This can be attributed to the fact that the repair operator reduces the diversity of the solutions. Therefore, we decided not to use the repair operator in the subsequent experiments.

We follow the published guidelines~\footnote{\url{https://www.obitko.com/tutorials/genetic- algorithms/recommendations.php}} used in prior work~\cite{DBLP:journals/tse/PanGB24} for setting GA hyperparameters. Specifically, we use a population size of $100$, a mutation rate of $0.01$, and a crossover rate of $0.90$. For each generation, the size of each solution (i.e., subset) remains fixed at the minimization budget.

\section{Study Design}\label{sec:studydesign}

\subsection{Research Questions}

\begin{enumerate}[label=RQ\arabic*:]
    \item How does \ourapproach~perform regarding TSM under different configurations?

    This RQ aims to identify the optimal configuration of \ourapproach~that achieves the best performance \textcolor{black}{in terms of \textit{FDR}}. Specifically, a \ourapproach~configuration entails the selection of a preprocessing method (Section~\ref{sec:preprocessing}), an embedding technique (Section~\ref{sec:representation}), a similarity measure (Section~\ref{sec:similarity}), and an initialization strategy (Section~\ref{sec:initialization}).
    
    \item How does \ourapproach~compare to baseline TSM techniques?

   This RQ aims to compare the performance of \ourapproach~with two baseline approaches, namely Random Minimization and FAST-R~\cite{cruciani2019scalable}, under various minimization budgets (i.e., 10\%, 20\%, ..., 90\%).

    \item How does the redundancy level of the test suite impact the effectiveness of TSM approaches?
    
    The performance of the TSM technique is influenced not only by the selected configuration but also by the characteristics of the test suites.
    This RQ aims to investigate and quantify the impact of the test suite redundancy level on the effectiveness (i.e., \textit{FDR}) of TSM.

    \item \textcolor{black}{How does \ourapproach~scale with test suite size compared to baseline TSM techniques?}

    \textcolor{black}{This RQ aims to assess the scalability of \ourapproach~compared to the baseline techniques. Specifically, we investigate how the run time--- including preparation time, search time and total minimization time---of \ourapproach~scale with test suite size, compared to baseline techniques.}

\end{enumerate}

\subsection{Dataset} \label{sec:dataset}

We evaluated \ourapproach~and compared it to baselines on a dataset collected from our industry partner, an automotive system company. There are $736$ test cases across seven test runs. 
As shown in Figure~\ref{fig:tc_example}, the test cases consist of test steps that validate the behavior of the automotive system related to a specific DTC. Each test case covers one requirement, covering a total of $54$ requirements, and may detect multiple faults, detecting a total of $220$ unique faults. 

Note that we identified the faults, which are the root causes of the failing test cases (e.g., parameter value mismatches, connection errors in specific components), using test execution logs under the guidance of the system engineers. For each test case, we first extracted failure messages from the test execution logs and then grouped those triggered by the same root cause into faults.

\noindent\textbf{\textit{Test suites for different redundancy levels. }}
In this dataset, a single fault can be detected by multiple test cases, indicating redundancy in the test suite. We define the redundancy level for the test suite as follows:

    \begin{equation}
    RL = \frac{\sum_{i = 1}^{m}t_{f_i}}{F_{unique}}
    \end{equation}
    where $m$ is the total number of test cases in this test suite, $t_{f_i}$ is the number of faults detected by the test case $t_i$, and $F_{unique}$ is the number of unique faults detected by this test suite.

For example, for a test suite $\mathcal{T}$ with three test cases that detect $3$, \textcolor{black}{$2$}, and $3$ faults, respectively. The total number of unique faults detected by $\mathcal{T}$ is $4$, the redundancy level of $\mathcal{T}$ is therefore \textcolor{black}{$(3 + 2 + 3)/4=2$}. Similarly, the redundancy level of our dataset is $2610/220 = 11.86$.

To assess the impact of the test suite redundancy level on the performance of \ourapproach, we generate $10$ diverse test suites for each redundancy level using Integer Linear Programming (ILP)~\footnote{https://pypi.org/project/PuLP/} and GA. This process starts by utilizing ILP to search for $100$ test suites with different test suite sizes while satisfying three constraints: (1) $100\%$ requirement coverage, (2) $100\%$ fault coverage, and (3) a specific test redundancy level. We set $15$ different redundancy levels, ranging from $4.5$ to $11.5$, with an interval of $0.5$. Note that no test suite satisfies a redundancy level below $4.5$ under constraints (1) and (2), and the highest redundancy level is $11.86$, which is observed in the original test suite. Then, to enhance the diversity of the dataset for experiments, for the $100$ test suites under each redundancy level, we employ GA to search for the most diverse $10$ test suites that minimize the sum of the set similarity between test suites, which is defined as the Jaccard Similarity between two test suites:

    \begin{equation}
    Jaccard~Similarity = \frac{|T_i \cap T_j|}{|T_i \cup T_j|}
    \end{equation}
    where $|T_i$ $\cap$ $T_j|$ denotes the number of test cases in the intersection of test suite $T_i$ and $T_j$. $|T_i \cup T_j|$ denotes the number of test cases in the union of test suites $T_i$ and $T_j$.

\subsection{Text Representations}\label{sec:representations}
Inspired by prior work on vector representations of requirements~\cite{zhao2025machine, abbas2025requirements} and test cases~\cite{viggiato2022identifying}, we employ a variety of text embedding techniques to convert test cases into vector representations. Specifically, we use TF-IDF, USE, Amazon Titan Text Embedding V2, and LongT5 to generate sentence-level embeddings, while Word2Vec, GloVe, and FastText are used to obtain word-level embeddings. Due to limitations in computational resources, data privacy constraints, and restrictions on input token length, we were unable to experiment with models such as BERT~\cite{devlin2019bert} and Sentence-BERT~\cite{DBLP:conf/emnlp/ReimersG19}.

\noindent\textbf{TF-IDF~\cite{salton1988term}.} TF-IDF has demonstrated promising performance in identifying similar test cases expressed in natural language~\cite{viggiato2022identifying}. We also employed TF-IDF to extract numerical vector representations of the test cases. For each word, we calculated its importance within a single test case relative to its occurrence across all other test cases.

\noindent\textbf{USE (Universal Sentence Encoder)~\cite{DBLP:journals/corr/abs-1803-11175}.} USE encodes text into high-dimensional vectors, capturing semantic information at the sentence level, making it particularly effective for tasks such as semantic similarity, text classification, clustering, and information retrieval~\cite{DBLP:journals/corr/abs-1803-11175}.

\noindent\textbf{LongT5~\cite{DBLP:conf/naacl/GuoAUONSY22}}. LongT5 is an extension of the T5 architecture designed to efficiently handle long sequences~\cite{DBLP:conf/naacl/GuoAUONSY22}. It incorporates sparse attention and reversible layers to reduce memory usage, making it well-suited for tasks like long-document summarization and retrieval-augmented generation. We selected the local attention mechanism for LongT5 because it yielded a higher \textit{FDR} for TSM than the global attention mechanism. 

\noindent\textbf{Amazon Titan Text Embedding V2~\cite{aws2024titantv2}.} 
For LLM-based embeddings, we use Amazon Titan Text Embedding V2, provided by our industrial partners. The model takes test case steps as input and produces a $1,024$-dimensional vector representation for test cases.

\noindent\textbf{Word2Vec~\cite{DBLP:journals/corr/abs-1301-3781}.} Word2Vec has been shown to be effective for identifying similar test cases~\cite{viggiato2022identifying}. In this study, we used the Continuous Bag-of-Words (CBOW) architecture with a window size of $10$ for Word2Vec, as this configuration yielded the highest \textit{FDR} for TSM. This was determined through experiments evaluating both the CBOW and Skip-Gram architectures with window sizes ranging from $2$ to $10$, in increments of $2$.
We trained Word2Vec models on the entire test case corpus and transformed each word into $300$-dimensional numerical vectors.

\noindent\textbf{GloVe~\cite{DBLP:conf/emnlp/PenningtonSM14}.} GloVe is an unsupervised learning algorithm that leverages global word-word co-occurrence statistics from a corpus to learn word embeddings~\cite{DBLP:conf/emnlp/PenningtonSM14}. Its core idea is that the ratio of co-occurrence probabilities can capture meaningful linear substructures within the word vector space. We selected a window size of $2$ for GloVe based on its performance in hyperparameter tuning experiments with window sizes ranging from $2$ to $10$, in increments of $2$. As in Word2Vec, we used GloVe to generate a $300$-dimensional numerical vector representation for each word in the test cases.

\noindent\textbf{FastText~\cite{DBLP:journals/tacl/BojanowskiGJM17}.} FastText is a word embedding model proposed by Bojanowski et al.~\cite{DBLP:journals/tacl/BojanowskiGJM17}, which extends Word2Vec by incorporating subword information. Each word is represented as a combination of character n-grams, enabling the model to better handle rare and out-of-vocabulary words, and to generalize more effectively—particularly in morphologically rich or noisy text. We used the CBOW architecture with a window size of $8$ as the FastText configuration, based on experiments evaluating both CBOW and Skip-Gram architectures with window sizes ranging from $2$ to $10$ in increments of $2$. We used FastText to convert each word in the test cases into a $300$-dimensional numerical vector.

\subsection{Baselines}
\label{sec:baselines}
\textcolor{black}{We compare \ourapproach~against (1) Random Minimization, a standard TSM baseline~\cite{pan2023atm}, (2) FAST-R approaches~\cite{cruciani2019scalable}, a set of novel TSM approaches with two versions: fixed-size budget version and adequate version. The former satisfies the pre-defined minimization budget constraint, whereas the latter satisfies the full requirement coverage constraint.}
\textcolor{black}{While being a widely used approach for TSM~\cite{khan2018systematic}, Greedy-based TSM approaches iteratively select the test cases that maximize the coverage until the full coverage constraint is satisfied~\cite{miranda2017scope,noemmer2019evaluation}. However, in our dataset, each test case covers only one requirement, rendering this option equivalent to Random Minimization with a requirement coverage constraint, as described below.}

\subsubsection{Random Minimization} Since the goal of TSM is to reduce the number of test cases and, consequently, lower testing costs, the most straightforward approach is to randomly remove test cases to meet the minimization budget~\cite{DBLP:journals/tse/AliBHP10}. We evaluate the performance of \ourapproach~against Random Minimization with and without the requirement coverage constraint, denoted as RM-Req and RM-NoReq, respectively.
For the version with the constraint, we applied the same selection strategy as GA initialization described in Section~\ref{sec:problem_formalization}. For the version without the constraint, we randomly select $k$ test cases from the test suite, ensuring that $k$ adheres to the minimization budget.

\subsubsection{FAST-R~\cite{cruciani2019scalable}} 
FAST-R is a family of similarity-based TSM approaches, including FAST++, FAST-CS, FAST-pw, and FAST-all. FAST++ and FAST-CS utilize k-means++ clustering~\cite{DBLP:conf/soda/ArthurV07} and constructed coresets~\cite{DBLP:conf/kdd/BachemL018}. In contrast, FAST-pw and FAST-all leverage minhashing and locality-sensitive hashing~\cite{DBLP:books/cu/LeskovecRU14}. FAST-R can be applied in two scenarios: the fixed-budget and adequate scenarios. The fixed-budget version of FAST-R minimizes the test suite within a predefined budget, while the adequate version prioritizes preserving the requirement coverage of the minimized test suite over adhering to the budget. One limitation of FAST-R is that it cannot simultaneously guarantee both a minimization budget and full requirement coverage.
We compare \ourapproach~against FAST-R under both scenarios where FAST-R is applicable (i.e., fixed-budget and adequate test suite minimization).
For FAST-R, we relied on the publicly available replication package\footnote{\url{https://github.com/ICSE19-FAST-R/FAST-R}} provided by its authors to ensure a fair and consistent comparison.

\subsection{Evaluation Metrics}

\noindent\textbf{\textit{Fault Detection Rate (\textit{FDR}).}} We use the \textit{FDR} to assess the effectiveness of \ourapproach, which is defined as follows:

    \begin{equation}
    FDR = \frac{F^{'}}{F}
    \end{equation}
    where $F$ is the total number of unique faults detected by the test suite before minimization, $F^{'}$ is the total number of unique faults detected by the test suite after minimization. \textit{FDR} ranges from 0 to 1, where 0 indicates that the minimized test suite detects no faults, while 1 signifies that it successfully detects all faults.

\noindent\textbf{\textit{Requirement Coverage.}} Although \ourapproach~ensures full coverage of all requirements, one of the selected baseline approaches, i.e., FAST-R, can not guarantee complete coverage when ensuring the minimization budget. To provide a more comprehensive comparison, we also evaluate the requirement coverage of the baseline approaches. This is calculated as the ratio of the number of requirements covered by the minimized test suite to the total number of original requirements:

\begin{equation}
    Coverage = \textcolor{black}{\frac{|\mathcal{R'}|}{|\mathcal{R}|}}
\end{equation}
where \textcolor{black}{\( \mathcal{R'} \)} represents the set of requirements covered by the minimized test suite \textcolor{black}{\( \mathcal{T'} \)}, and \textcolor{black}{\( \mathcal{R} \)} denotes the original set of requirements. The coverage value ranges from 0 to 1, where 0 indicates that the minimized test suite covers no requirements, while 1 signifies full coverage of all requirements.

\subsection{Experiment Environment}
All experiments were conducted on a computer provided by our industry partner, running Windows 10, equipped with an Intel Core i7-11850H CPU at 2.5 GHz and 32 GB of RAM. Due to data privacy considerations, we were restricted to performing our experiments solely on this laptop, limiting the text representation techniques available to us.

\section{Results}\label{sec:results}

\subsection{Performance of \ourapproach~under different configurations (RQ1)}
\noindent \textbf{Approach.} To address RQ1, we systematically evaluated the performance of \ourapproach~under various configurations. Specifically, under a mid-range minimization budget ($50\%$), we ran \ourapproach~with three different preprocessing methods (as described in Section~\ref{sec:preprocessing}), and applied both word-level and sentence-level embedding techniques to transform test cases into vector representations (Section~\ref{sec:representation}).
To measure test case similarity, we compared several widely used metrics, including cosine similarity, Euclidean distance, and WMD. In addition, we investigated the impact of various GA initialization strategies on the overall performance of \ourapproach~by evaluating all three strategies introduced in Section~\ref{sec:initialization}.

Through a comprehensive grid search across these configurations, we identified the optimal configuration of \ourapproach, i.e., the combination of preprocessing method, embedding technique, similarity metric, and initialization strategy that yields the best performance in terms of \textit{FDR}. 

\noindent \textbf{Results.} Table~\ref{tab:rq1} presents the \textit{FDR} of \ourapproach~across different configurations under 50\% minimization budget. For each combination of initialization strategy and embedding technique, the best \textit{FDR} is shown in bold, while the overall best result for each embedding technique is highlighted with a gray background.

Overall, we observe that sentence-level embeddings generally outperform word-level embeddings in terms of \textit{FDR}. Among the sentence-level embedding techniques, TF-IDF combined with cosine similarity achieves the best overall performance, with a peak \textit{FDR} of $86.09\%$ under $PM3$ and \textit{Init Strategy 2}. 
Similarly, Amazon Titan Text Embedding V2, LongT5, and USE also achieve competitive results, with the best Titan configuration reaching $81.59\%$ ($PM3$, Init Strategy 2, Euclidean distance), LongT5 achieving $78.50\%$ ($PM3$, Init Strategy 3, Euclidean distance), and USE achieving $78.05\%$ ($PM1$, Init Strategy 3, Euclidean distance).

\nff{Interestingly, despite being a simple, sparse representation, TF-IDF outperformed all other sentence-level embeddings. 
The characteristics of our industrial test cases can explain this. As shown in Figure~\ref{fig:tc_example}, the test cases in our dataset follow \textit{structured} and \textit{repetitive} templates with little linguistic variation. The steps in the test cases differ only in variable names or parameter values, and TF-IDF precisely captures these differences through its word-frequency mechanism, down-weighting common words that appear in every test case and up-weighting rare words, including distinct variable names and parameter values.
In contrast, other sentence-level embeddings, while capable of capturing deep semantic relationships, 
tend to smooth out subtle lexical differences that are crucial for distinguishing test cases in our dataset.
This suggests that for template-based, domain-specific test descriptions, TF–IDF, which suppresses common terms and emphasizes rare tokens, often outperforms semantically rich embeddings that smooth over token-level distinctions.}

For word-level embeddings, Word2Vec slightly outperforms FastText and GloVe, but all three show lower \textit{FDR} compared to sentence-level embedding techniques. Their performance peaks are in the range of $75\%-76\%$, with Word2Vec achieving the highest result of $76.27\%$ under $PM1$ and Init Strategy 2.

Among similarity measures, Euclidean distance outperforms cosine similarity for sentence-level embeddings most of the time, suggesting that the absolute distance measured by Euclidean-based spatial separation is more effective for capturing test case diversity in the embedding space.

In terms of preprocessing, we observe that, for word-level embedding, $PM1$ and $PM2$ often yield better \textit{FDR} results than $PM3$, indicating that preprocessing is required for using word-level embedding techniques. Word-level embeddings are more sensitive to individual characters, such as casing and punctuation, so proper preprocessing helps normalize these variations, enabling more accurate similarity computations. On the other hand, $PM1$ and $PM3$ tend to lead to higher \textit{FDR} than using $PM2$ for sentence-level embeddings, meaning that less or no preprocessing can help these models better learn context information from test cases, without losing the key details that distinguish them.

Lastly, Initialization Strategies 2 and 3 generally lead to higher \textit{FDR} compared to Initialization Strategy 1, particularly when combined with stronger embeddings and similarity measures. This confirms the importance of starting the evolutionary search with a well-distributed and diverse initial subset.

\textcolor{black}{Note that we further conducted experiments under 25\% and 75\% minimization budgets. The results show that across these three minimization budgets, TF-IDF consistently yields the highest \textit{FDR}. While the best preprocessing method, initialization strategy, and similarity measure could differ slightly across budgets, the corresponding \textit{FDR} differences were within $0.025$ and do not affect the conclusions discussed above. This indicates that our configuration choice is not sensitive to the specific budget. Given that the $50\%$ minimization budget is a representative mid-range budget, we report the best configuration determined at $50\%$ and use it in subsequent comparisons. The results for other budgets are available in the replication package.}

\answer{
\textit{\textbf{Answering RQ1: TF-IDF with cosine similarity yields the highest overall \textit{FDR} (86.09\%) under Preprocessing Method PM3 and Initialization Strategy 2 at the 50\% minimization budget.} }}

\begin{table*}[!htbp]
\caption{\textit{FDR} (\%) of \ourapproach~across different configurations under 50\% minimization budget.}
\label{tab:rq1}
\centering
\resizebox{0.9\textwidth}{!}{
\begin{tabular}{c|c|ccc|ccc|ccc}
\hline
\multicolumn{1}{l|}{\multirow{2}{*}{}}         & Similarity & \multicolumn{3}{c|}{Init Strategy 1}      & \multicolumn{3}{c|}{Init Strategy 2}      & \multicolumn{3}{c}{Init Strategy 3}       \\ \cline{3-11} 
\multicolumn{1}{l|}{}  &  Measure  & $PM1$ & $PM2$ & $PM3$ & $PM1$ & $PM2$ & $PM3$ & $PM1$ & $PM2$ & $PM3$ \\ \hline
\multirow{2}{*}{TF-IDF} & Cosine & 80.59 & 80.32 & 80.41 & 84.36 & 83.00 & \cellcolor{gray!20} \textbf{86.09} & \textbf{85.41} & 81.91 & 83.09 \\
   & Euclidean & 82.32 & 81.14 & \textbf{83.27} & 84.91 & 83.68 & 85.14 & 85.05 & 84.41 & 85.18 \\ \hline
\multirow{2}{*}{USE}   & Cosine & 70.36 & 72.68 & 70.41 & 77.05 & 76.27 & 77.41 & 77.59 & 77.05 & 77.59 \\
   & Euclidean & 71.68 & \textbf{73.91} & 72.68 & \textbf{77.68} & 76.50 & 77.05 & \cellcolor{gray!20} \textbf{78.05} & 75.82 & 77.50 \\ \hline
\multirow{2}{*}{LongT5}& Cosine & 72.82 & 72.68 & 72.86 & 76.14 & 77.09 & 76.14 & 77.23 & 76.45 & 76.18 \\
   & Euclidean & 74.05 & 74.64 & \textbf{74.91} & 77.77 & 78.23 & \textbf{78.32} & 78.36 & 77.82 & \cellcolor{gray!20} \textbf{78.50} \\\hline
\multirow{2}{*}{\shortstack{Amazon Titan Text\\Embedding V2}} & Cosine & 76.55 & 72.00 & 76.05 & 81.50 & 74.77 & 80.86 & 79.05 & 75.82 & 79.18 \\
    & Euclidean & \textbf{78.45} & 72.27 & 77.95 & 80.82 & 75.86 & \cellcolor{gray!20} \textbf{81.59} & \textbf{79.91} & 75.59 & 79.86 \\ \hline
Word2Vec     & WMD & 69.36 & \textbf{71.23} & 68.45 & \cellcolor{gray!20} \textbf{76.27} & 74.00 & 75.00 & 75.68 & 74.27 & \textbf{75.82} \\\hline
GloVe        & WMD & 65.64 & \textbf{71.00} & 66.18 & 73.59 & \cellcolor{gray!20} \textbf{75.50} & 74.41 & 74.09 & \textbf{75.00} & 72.55 \\ \hline
FastText     & WMD  & 66.55 & \textbf{71.41} & 67.00 & \textbf{74.82} & 73.82 & 74.32 & 74.09 &\cellcolor{gray!20} \textbf{75.68} & 73.32\\ \hline
\end{tabular}
}
\end{table*}

\subsection{Performance of \ourapproach~compared with baselines (RQ2)}
\noindent \textbf{Approach.} To address RQ2, we compared the performance of \ourapproach~with the best configuration against baseline approaches in terms of \textit{FDR}, varying the minimization budget from $10\%$ to $90\%$ in $10\%$ increments. According to RQ1, we selected \ourapproach~using $PM3$ as preprocessing method, TF-IDF as the embedding technique, Initialization Strategy 2 for GA initialization, and cosine similarity for the similarity measure as the best-performing configuration.

We evaluated \ourapproach~against Random Minimization, which also employs Initialization Strategy 2, two versions of
FAST-R~\cite{cruciani2019scalable}—the fixed-budget and adequate versions.
For the fixed-budget version of FAST-R, we varied the minimization budget from $10\%$ to $90\%$, in increments of $10\%$. \textcolor{black}{The fixed-budget version of FAST-R and RM-NoReq do not guarantee the 100\% requirement coverage and are included in this RQ solely for \textit{FDR} comparison.} In the adequate scenario, we first ran FAST-R to determine the resulting minimization budget. This budget was then applied consistently to \ourapproach~and other baselines for experimental comparison. Notably, due to redundancy in our dataset, as few as $7\%$ (i.e., $54$ test cases, one test case per requirement) are sufficient to achieve $100\%$ requirement coverage. 
 
Moreover, for comparison purposes, we calculated the theoretical best possible \textit{FDR} for the test suite under different minimization budgets using an ILP approach. Specifically, we employed ILP to find the subset that (1) satisfies the minimization budget, (2) achieves $100\%$ requirement coverage, and (3) maximizes the number of covered faults.

\begin{figure}[!htbp]
\centering
\includegraphics[width=0.6\linewidth]{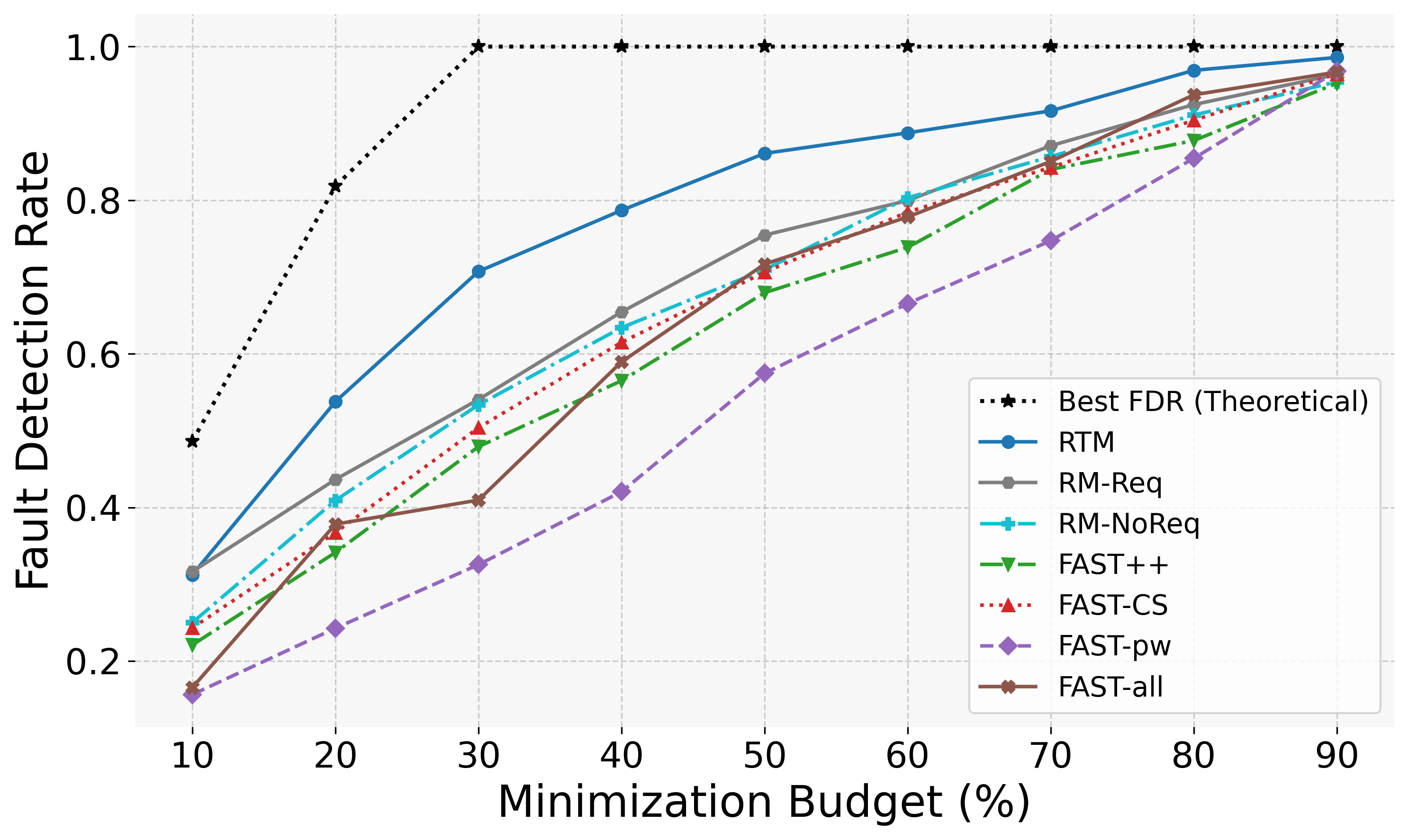}
\Description{}
\caption{Comparison of \textit{FDR} across minimization budgets for \ourapproach, FAST-R, and Random Minimization, alongside the theoretical upper bound.}
\label{fig:rq2}
\end{figure}

\noindent \textbf{Results.} Figure~\ref{fig:rq2} depicts the \textit{FDR} for \ourapproach, Random Minimization, the fixed-budget version of FAST-R, 
as well as the theoretical best possible \textit{FDR} across minimization budgets from $10\%$ to $90\%$.

We observe that \ourapproach~consistently outperforms Random Minimization and FAST-R approaches in terms of \textit{FDR} across various minimization budgets. The performance gap is particularly notable at mid-range budgets ($30\%$ to $60\%$). For instance, at a $40\%$ minimization budget, \ourapproach~achieves an \textit{FDR} of $78.68\%$, whereas RM-Req and FAST++ yield approximately $65.45\%$ and $56.50\%$, respectively. This demonstrates \ourapproach’s superior ability to retain fault-revealing test cases, especially under mid-range budget constraints. At lower and higher minimization budgets, the \textit{FDR} of all techniques tends to converge toward either $0$ or $1$, thus making the difference in \textit{FDR} between \ourapproach~and other techniques smaller. However, mid-range minimization budgets are more important in practice, as they offer more useful trade-offs between cost savings and fault detection.

As the minimization budget increases, the \textit{FDR} of all techniques gradually improves and approaches the theoretical upper bound of $1.0$. Among the FAST-R variants, FAST-CS and FAST-all demonstrate better performance, while FAST++ and FAST-pw exhibit weaker results at lower budgets, indicating limited effectiveness in resource-constrained settings. Interestingly, Random Minimization generally achieves higher \textit{FDR} than FAST-R. This may be due to the distance metrics employed by FAST-R, which may not provide effective guidance for reducing similar test cases, whereas Random Minimization inherently preserves a greater diversity of test cases, leading to better \textit{FDR}. \textcolor{black}{Moreover, the fixed-budget version of FAST-R does not guarantee full requirement coverage, making it unusable in our context.}

For the adequate version of FAST-R, after reducing similarity among test cases while preserving requirement coverage of the minimized test suite,
FAST++, FAST-CS, and FAST-pw ultimately retained only $7\%$ (i.e., selecting one test case under each requirement) of the test cases, while FAST-all retained $90\%$. We therefore compared \ourapproach~
and Random Minimization using $7\%$ and $90\%$ as the minimization budgets with the adequate version of FAST-R.
The detailed results are presented in Table~\ref{tab:rq2}. Under the $7\%$ budget, \ourapproach~achieves a high \textit{FDR} of $26.32\%$ while maintaining full requirement coverage ($100\%$), which is comparable to FAST++ ($27.27\%$) and FAST-CS ($25.95\%$), and outperforming FAST-pw in both \textit{FDR} ($21.18\%$) and requirement coverage ($89.44\%$). RM-Req and RM-NoReq achieve \textit{FDR} of $25.91\%$ and $19.86\%$, respectively. 

Under the $90\%$ minimization budget, \ourapproach~achieves the highest \textit{FDR} of $98.59\%$ among all techniques, 
maintaining full requirement coverage. In contrast, FAST-all shows slightly lower \textit{FDR} ($92.59\%$) and slightly reduced coverage ($99.63\%$). RM-Req and RM-NoReq yield lower \textit{FDR} of $96.36\%$ and $95.41\%$, respectively. 

\textcolor{black}{These results demonstrate that, \ourapproach~consistently delivers competitive \textit{FDR} without compromising requirement coverage in the adequate scenario.}

\begin{table}[htbp]
\centering
\caption{\textit{FDR} (\%) and requirement coverage (\%) comparison between \ourapproach~and baselines under the adequate-scenario setting.}\label{tab:rq2}
\renewcommand{\arraystretch}{1.2}
\resizebox{0.6\linewidth}{!}{
\begin{tabular}{l|ccc|ccc}
\hline
\multirow{2}{*}{Approach} & \multicolumn{3}{c|}{\textbf{7\% Budget}} & \multicolumn{3}{c}{\textbf{90\% Budget}} \\
                        & \textit{FDR}   & Coverage &           & \textit{FDR}   & Coverage &           \\ \hline
\ourapproach                    & 26.32  & 100.00    &           & 98.59  & 100.00    &           \\
FAST++                 & 27.27  & 100.00    &           & ---  & --     &           \\
FAST-CS                & 25.95  & 100.00    &           & --    & --       &           \\
FAST-pw                & 21.18  & 89.44     &           & --    & --       &           \\
FAST-all               & --    & --       &           & 92.59  & 99.63     &           \\
RM-Req               & 25.91    & 100.00       &           & 96.36    & 100.00       &           \\ 
RM-NoReq                & 19.86    & 39.26       &           & 95.41    & 98.52       &           \\ 
\hline
\end{tabular}}
\end{table}

\textcolor{black}{In summary, compared with the baselines, \ourapproach~achieves the highest \textit{FDR} on most minimization budgets, making it a
more reliable choice for practical TSM with strong fault detection guarantees.}
In addition, \ourapproach~ensures full requirement coverage while enabling pre-defined minimization budgets, two features not simultaneously supported by FAST-R. 
These capabilities are particularly crucial in the context of requirement-based testing, where covering all requirements is often a strict necessity. Moreover, the ability to ensure fixed-budget minimization is essential in resource-constrained environments, where testing time or execution cost must be carefully managed.

\answer{
\textcolor{black}{\textit{\textbf{Answering RQ2: \ourapproach~consistently outperforms all baselines in both \textit{FDR} and requirement coverage on most minimization budgets.}}}}

\subsection{Impact of test suites redundancy (RQ3)}

\noindent \textbf{Approach.}
In RQ3, we investigate the impact of test suite redundancy levels on the performance (i.e.,\textit{FDR}) of TSM techniques. Evaluating TSM techniques necessarily requires considering the test suite redundancy level, and we aim to shed light on this aspect. As described in Section~\ref{sec:dataset}, for each redundancy level (ranging from $4.5$ to $11.5$ in increments of $0.5$), we employ the ILP algorithm and GA to generate $10$ diverse test suites, each ensuring full coverage of both requirements and faults.
We then apply \ourapproach~and the baseline techniques (Random Minimization and fixed-budget version of FAST-R) to each of these test suites, and report the average \textit{FDR} across the $10$ test suites for each redundancy level. \textcolor{black}{Note that the fixed-budget version of FAST-R and RM-NoReq do not guarantee 100\% requirement coverage. We include them in this RQ solely for \textit{FDR} comparison.}

\begin{figure*}[htbp!]
\centering
\includegraphics[width=\linewidth]{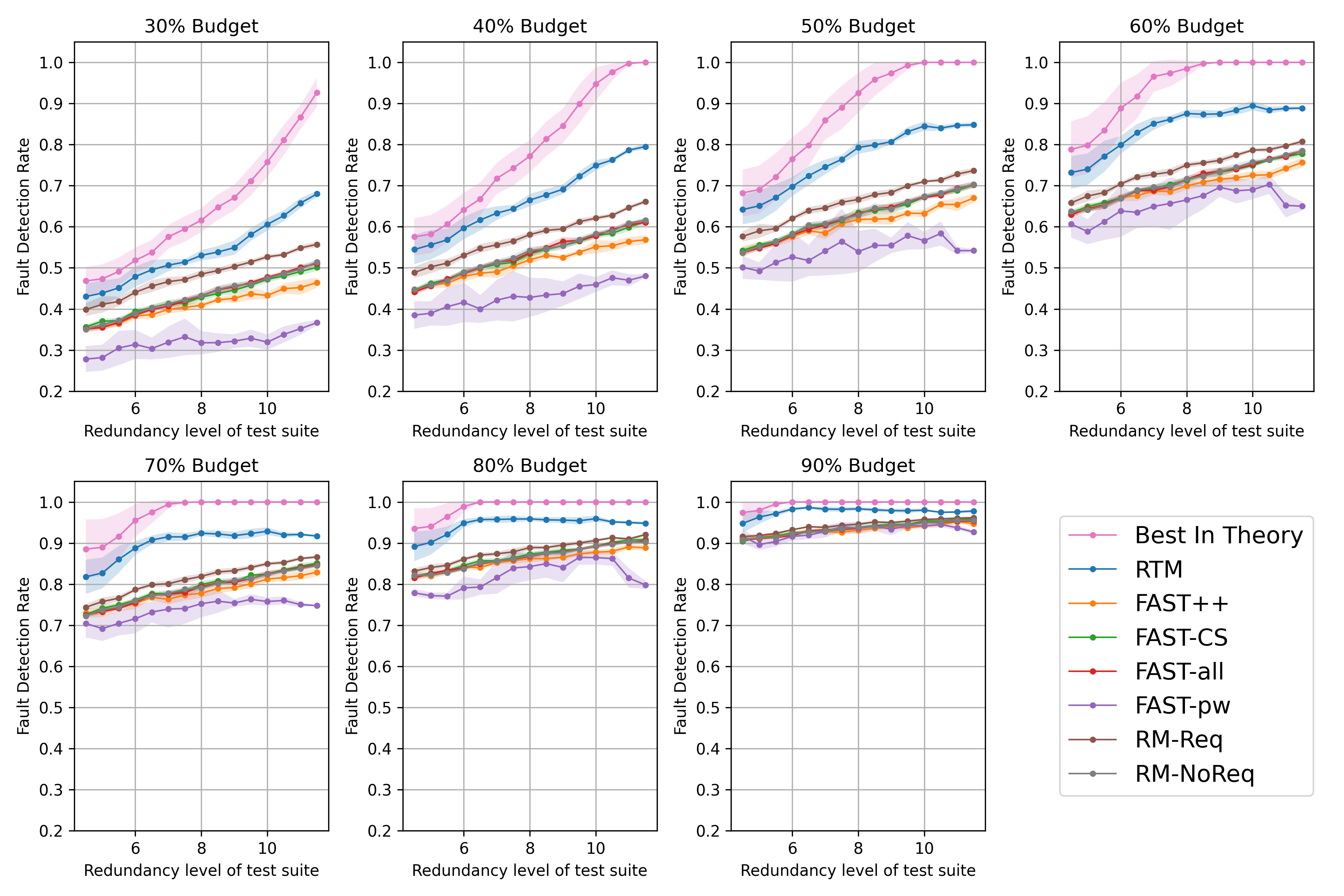}
\Description{}
\caption{Comparison of \textit{FDR} across varying redundancy levels for \ourapproach~and baseline approaches under seven minimization budgets.}
\label{fig:rq3}
\end{figure*}

\noindent \textbf{Results.} 
Figure~\ref{fig:rq3} illustrates the \textit{FDR} performance of \ourapproach~against baselines across varying redundancy levels (ranging from $4.5$ to $11.5$ in increments of $0.5$) under seven distinct minimization budgets (from $30\%$ to $90\%$ in increments of $10\%$). Moreover, for each test suite, we calculate the theoretical best \textit{FDR} for comparison purposes. Note that we cannot set minimization budgets below $30\%$ because some test suites cannot achieve full requirement coverage under those constraints.

We observe that \ourapproach~consistently achieves superior \textit{FDR} compared to all baseline techniques across all redundancy
levels and minimization budgets, while aligning more closely with the theoretical optimum. The difference in terms of \textit{FDR} between \ourapproach~and baselines is particularly significant
at mid-range minimization budgets (30\%–60\%).

As expected, the \textit{FDR} of all techniques improves as the test suite redundancy level increases, indicating that the TSM techniques tend to perform better--achieving higher \textit{FDR}--when applied to more redundant test suites. This trend, however, becomes less evident at higher budgets ($80\%$ and $90\%$), where the \textit{FDR} of all techniques converge toward $1$. 
Random Minimization (RM-Req and RM-NoReq) slightly outperforms other FAST-R variants but consistently falls short of \ourapproach. 
We also observed that across all minimization budgets and redundancy levels, RM-Req consistently outperformed RM-NoReq in \textit{FDR}. This demonstrates that, for our dataset, explicitly ensuring that every requirement is covered not only preserves full requirement coverage but also significantly boosts the test suite's ability to detect faults. In other words, prioritizing requirement coverage is an effective heuristic for improving fault detection under budgetary constraints.

Moreover, the analysis reveals that redundancy level significantly influences \textit{FDR} at lower minimization budgets (below $60\%$), highlighting the importance of considering redundancy under tight budget constraints. Notably, higher redundancy combined with low minimization budgets tends to result in larger gaps between the \textit{FDR} of all TSM techniques and the theoretical maximum. This discrepancy might be explained as follows: under a very low minimization budget, the search space is severely restricted, making it challenging to assemble a diverse subset that both covers all requirements and maximizes fault coverage. Future research should account for redundancy levels when evaluating TSM techniques.

\answer{
\textit{\textbf{Answering RQ3: 
The test suite redundancy level generally has a strong effect on the achieved \textit{FDR} of all TSM techniques. 
Further, \ourapproach~consistently outperforms all baseline techniques in terms of \textit{FDR} across all redundancy levels and minimization budgets.}}}

\textcolor{black}{\subsection{Scalability of \ourapproach~compared with baselines (RQ4)}
\label{sec:rq4}}

\textcolor{black}{\noindent \textbf{Approach.} To address RQ4, we compared the performance of \ourapproach~with the best configuration against baselines in terms of scalability, under $50\%$ minimization budget. As described in Section~\ref{sec:dataset}, we generate test suites under $15$ redundancy levels with various test suite sizes, resulting in a total of $150$ test suites. To investigate how \ourapproach~scales with test suite size compared to baselines, we collect the runtime, including preparation time (time for embeddings generation and similarity calculation), search time (time for searching the optimal subsets), and total minimization time (the summation of preparation time and search time), across $150$ test suites with different test suite sizes. Moreover, to quantify and compare the relationships between runtime and test suite size across different approaches, we fit regression models for preparation time, search time, and total minimization time as functions of test suite size. }

\begin{figure*}[htbp!]
\centering
\includegraphics[width=\linewidth]{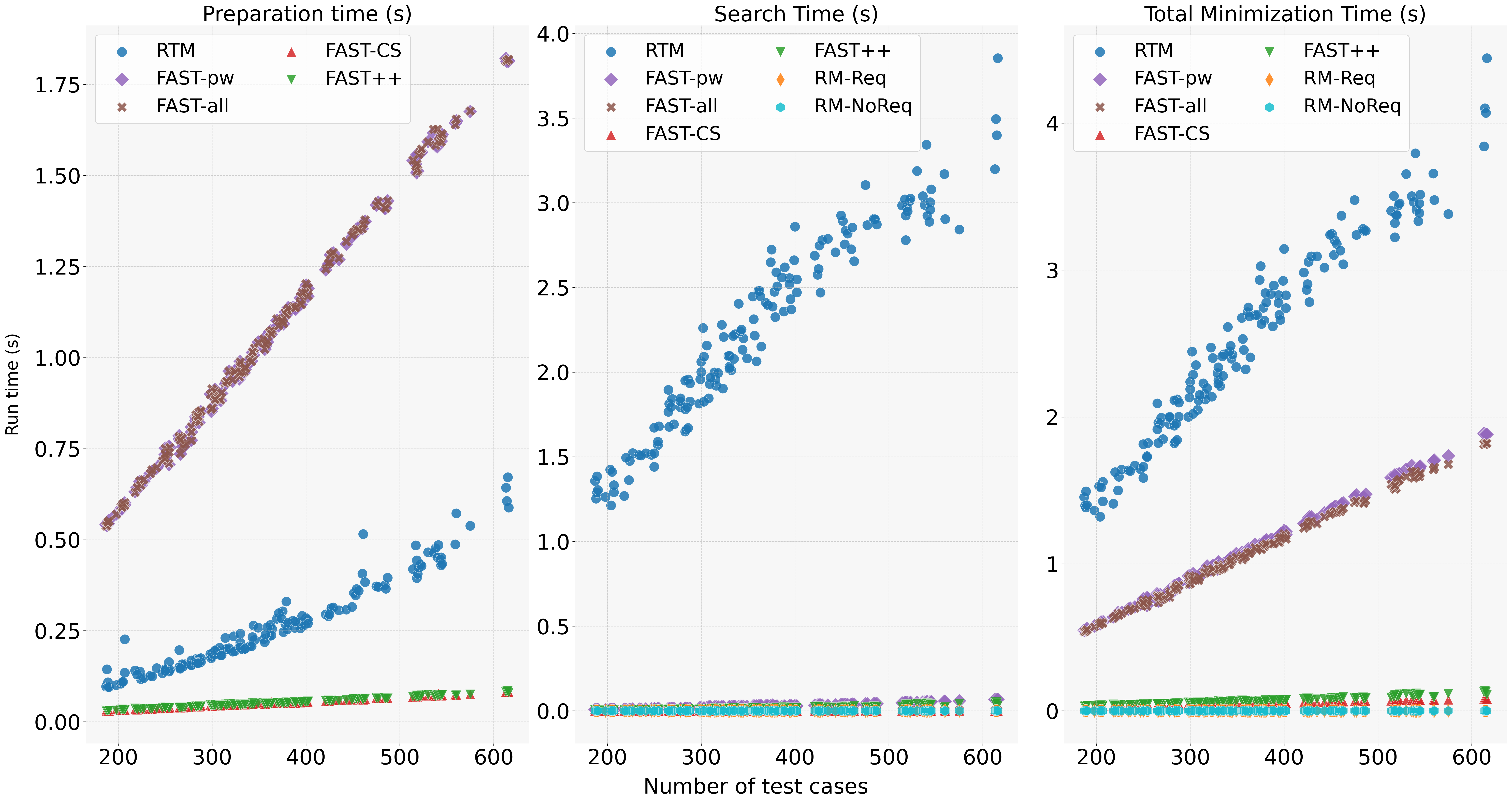}
\Description{}
\caption{Scatter plots of the number of test cases and preparation time, search time and total minimization time (in sec) for \ourapproach~and baselines under 50\% minimization budget.}
\label{fig:rq4}
\end{figure*}

\textcolor{black}{\noindent \textbf{Results.}
Figure~\ref{fig:rq4} depicts how the preparation time, search time, and total minimization time of \ourapproach~and baseline approaches scale with the test suite sizes (number of test cases) under a $50\%$ minimization budget. We observe that the observations in Figure~\ref{fig:rq4} approximately follow a linear relationship. Therefore, for each approach, a linear regression model is fitted to the preparation time, search time, and total minimization time as functions of test suite size. The equation of the linear regression model is as follows: }

{\color{black}
    \begin{equation}
    Time = a * n + c 
    \end{equation}
}
\textcolor{black}{where $n$ denotes the number of test cases, $a$ is the regression coefficient of the linear term, $c$ denotes the intercept of the model.}

\textcolor{black}{The results of the linear regression model confirm that the total minimization time, preparation time, as well as search time and the number of test cases, for \ourapproach~and baseline approaches, follow a linear relationship with the coefficient on the linear term statistically significant with $\alpha = 0.01$. The $R^2$ values for all the regression models are above $93\%$, indicating a high percentage of variation in runtime can be explained by the number of test cases.}

\textcolor{black}{We observe that, although being slower than all other baselines, \ourapproach~minimizes test cases in seconds--$1.32$s to $4.44$s in terms of total minimization time--for test suite size from $187$ to $616$, which is computationally efficient. Moreover, the fact that the total minimization time of \ourapproach~follows a linear relationship with the test suite size indicates that \ourapproach~is scalable for larger test suites. According to the fitted linear regression model, when the test suite size $n$ is $200{,}000$, the predicted total minimization time of \ourapproach~is 
$6.217686\times10^{-3}\,n + 2.523669 \approx 1246.06$ seconds, whereas FAST-CS (the fastest FAST-R approach) is $1.195291\times10^{-4}\,n + 0.051979 \approx 23.96$ seconds and
RM-Req is $9.062634\times10^{-7}* n +  0.000472 \approx 0.18$ seconds. Thus, the time differences are $20.37$ mins and $20.76$ mins when compared to FAST-CS and RM-Req, respectively. Given the fact that, unlike test case selection and prioritization, test suite minimization only performs occasionally, such as major releases with many new added test cases and redundant old test cases~\cite{DBLP:journals/tse/PanGB24,pan2023atm}, such differences are practically acceptable.}

\textcolor{black}{Moreover, the preparation time of \ourapproach~ranges from $0.10$s to $0.67$s, which is negligible in our context. We also observe that \ourapproach~is much faster than FAST-all in terms of preparation time. According to the regression model results, when the test suite size is $200{,}000$, the predicted absolute preparation time difference between \ourapproach~and FAST-CS is $3.30$ min, which is of no practical consequence for TSM. As for the search time, \ourapproach~took $1.21$s to $3.85$s across $150$ test suites. When the test suite size is  $200{,}000$, the predicted absolute search time difference of \ourapproach~ when compared to FAST-CS is $17.07$ mins.}

\textcolor{black}{In summary, \ourapproach~runs within $5$ seconds across all evaluated test suites. Although being slower than the baselines, \ourapproach~scales approximately linearly with test suite sizes, and even for substantially larger suites, the absolute runtime gaps remain acceptable in practice---they are not expected to have practical consequences for TSM. Given that \ourapproach~consistently achieves much higher \textit{FDR} across various datasets and minimization budgets than the baselines while preserving full requirement coverage, it is therefore a better option in many practical contexts.}

\answer{
\textit{\textbf{\textcolor{black}{Answering RQ4: \ourapproach~is scalable in terms of runtime and, on the evaluated datasets, exhibits small absolute time differences (i.e., of no practical consequences) compared to the baselines while achieving much higher \textit{FDR} performance with full requirement coverage preserved, thus making it a better choice in practice. } }}}

\section{Threats to Validity}\label{sec:threats}
In this study, we identified the following threats to validity:

\noindent
\textbf{External Validity.}
The generalizability of our findings may be limited due to the specific domain and nature of the test suites provided by our industry partner. The performance of our approach might differ when applied to other domains or software systems with different requirement structures, test suite characteristics, or fault profiles. To mitigate this threat, we selected test suites of varying sizes and redundancy levels to evaluate the robustness of our approach across diverse configurations. Furthermore, our approach is domain-agnostic, relying solely on textual information from requirements and test cases, thereby facilitating adaptation to other domains. Lastly, our industry partner follows standard development practices in the automotive domain, as defined by standards similar to those in other safety-critical domains where functional safety is important. Future work includes evaluating our approach on publicly available datasets from different domains to further assess the generalizability.

\noindent
\textbf{Internal Validity.}
To ensure a fair and consistent comparison, we adopt the original replication package of FAST-R provided by its authors. This minimizes potential implementation biases or configuration inconsistencies that could otherwise affect the validity of comparative results.

Another threat to external validity arises from the limited choice of embedding techniques, constrained by the data privacy requirements of our industry partner. This limitation may have affected the overall performance of our approach. Nevertheless, to mitigate this threat, we evaluated seven different embedding methods within the bounds of our available computational resources. This broad comparison helps reduce concerns regarding internal validity, despite the constraints on embedding selection.

\noindent
\textbf{Construct Threats.}
\nff{A potential construct threat concerns the maintenance cost of TF-IDF representations when new test cases are added. In our context, this risk is limited because the dataset originates from automotive system requirement-based test cases, which follow well-defined structural templates and rely on a stable, domain-specific vocabulary. As a result, the introduction of new terms is infrequent and typically occurs only during major version updates, making the need for corpus-wide re-computation of TF-IDF vectors relatively unlikely. Moreover, as discussed in Section~\ref{sec:rq4}, the preparation time of TF-IDF is scalable for large test suites, and the maintenance cost is therefore acceptable.
However, in domains where the vocabulary evolves rapidly or new linguistic expressions appear frequently (e.g., user-generated content or dynamic project documentation), we recommend using domain-specific or LLM-based embeddings that can better adapt to incremental updates without requiring full re-vectorization.
}

\noindent
\textbf{Conclusion Threats.}
To reduce the impact of randomness and increase the reliability of our results, we run each experiment ten times with different random seeds and report the average performance. This mitigates the risk of drawing conclusions based on outlier results or fluctuations arising from stochastic components in the minimization process. 

\section{Related Work}
\label{sec:relatedwork}
Several approaches have been proposed to support TSM, which can be classified in to three categories~\cite{khan2018systematic, cruciani2019scalable, pan2023atm}: greedy~\cite{noemmer2019evaluation, miranda2017scope}, clustering~\cite{coviello2018clustering, liu2011user, viggiato2022identifying, cruciani2019scalable}, and search-based~\cite{zhang2019uncertainty, hemmati2013achieving, pan2023atm, DBLP:journals/tse/PanGB24, marchetto2017combining}, plus hybrid combinations~\cite{xia2021test, anwar2019hybrid, yoo2010using} thereof.

\noindent
\textbf{Greedy-based approaches} follow a heuristic, step-by-step selection strategy, where the test case that provides the most benefit toward the objective is chosen at each iteration, until the given constraints (e.g., full requirement coverage or code path coverage) are satisfied. Miranda et al.~\cite{miranda2017scope} and Noemmer et al.~\cite{noemmer2019evaluation} both employed greedy heuristics for TSM based on statement coverage. Their approaches iteratively selected test cases to maximize code coverage, achieving reasonable \textit{FDR}s while significantly reducing the number of executed test cases. Greedy-based approaches are computationally efficient and easy to implement, but may suffer from suboptimal global performance due to their locally optimal decisions. \textcolor{black}{
Moreover, given that, in our dataset, each test case covers only one requirement, the coverage-based greedy strategy--iteratively selecting the test cases that maximize the coverage until the full coverage constraint is satisfied--is equivalent to Random Minimization with the requirement constraint (RM-Req), as described in Section~\ref{sec:baselines}.}

\noindent
\textbf{Clustering-based approaches} group similar test cases based on syntactic or semantic similarities, such as those derived from natural language descriptions or execution behavior. The goal is to reduce redundancy by selecting representative test cases from each cluster. These approaches are particularly useful in settings where test cases are written in natural language and code coverage information is unavailable. Viggiato et al.~\cite{viggiato2022identifying} proposed a comprehensive approach that combines text embedding, similarity measurement, and clustering to identify and remove similar test cases written in natural language. They split the requirement test cases into test steps and then calculate the similarity between test steps. Cruciani et al.~\cite{cruciani2019scalable} proposed FAST-R, a black-box TSM approach that leverages the source code of test cases without requiring execution or coverage information. FAST-R transforms test code into vector representations using a term frequency model, followed by dimensionality reduction via random projection. Similarly, Coviello et al.~\cite{coviello2018clustering} proposed a clustering-based approach using Hierarchical Agglomerative Clustering to perform TSM.

\noindent
\textbf{Search-based approaches} formulate the minimization problem as an optimization task and utilize metaheuristic algorithms (e.g., GA, simulated annealing) to explore the space of possible test subsets. These methods can balance multiple conflicting objectives, such as minimizing the number of test cases while maximizing fault detection capability and maintaining requirement coverage. Hemmati et al.~\cite{hemmati2013achieving} applied a search algorithm that minimizes test suites based on model-derived test case similarity. Similarly, Zhang et al.\cite{zhang2019uncertainty} proposed UncerTest, a model-based test case minimization framework that leverages multi-objective search algorithms, including NSGA-II, MOCell\cite{nebro2009mocell}, and SPEA2~\cite{zitzler2001spea2}. ATM~\cite{pan2023atm} is a black-box TSM approach that combines test code similarity analysis with evolutionary search. It represents test code using Abstract Syntax Trees and employs GA and Non-Dominated Sorting Genetic Algorithm II~\cite{meyarivan2002fast} as the underlying evolutionary algorithms. Furthermore, LTM~\cite{DBLP:journals/tse/PanGB24} leverages LLMs (CodeBERT, UniXcoder, and CodeLlama) to compute similarity between test cases and employs a GA to search for an optimal subset of test cases under a fixed budget. 
MORE+~\cite{marchetto2017combining} is a multi-objective TSM approach that uses NSGA-II to optimize structural, functional, and cost-related objectives, aiming to enhance fault detection while reducing redundancy and execution time.
Although typically more computationally intensive than greedy and clustering approaches, search-based techniques support multiple minimization objectives and often achieve better performance for TSM.

\noindent
\textbf{Hybrid approaches} integrate complementary methods into a unified framework to balance multiple objectives, including reducing test suite size, preserving fault detection capability, and satisfying requirement or coverage constraints. Xia et al.~\cite{xia2021test} apply clustering-based test suite reduction combined with evolutionary multi-objective optimization. Anwar et al.~\cite{anwar2019hybrid} combined a GA and particle swarm optimization to optimize regression test suites. Yoo et al.~\cite{yoo2010using} employ a hybrid multi-objective GA that integrates the efficient approximation of the greedy approach with the population-based GA's ability to generate higher-quality Pareto fronts.

While promising, existing techniques primarily focus on code-based solutions for minimization. In contrast, our work addresses minimization in the context of requirement-driven testing, where test cases are specified in natural language, which introduces distinct challenges. Moreover, current approaches do not explicitly support TSM under a fixed budget while simultaneously enforcing full requirement coverage as a hard constraint. \textcolor{black}{To this end, we adopt a GA, as in the SOTA TSM work (i.e., ATM and LTM), and augment it with an explicit requirement coverage constraint that guarantees $100\%$ coverage under a fixed minimization budget. Note that we did not use NSGA-II, as reported in the ATM work, since it achieves comparable performance while being substantially more time-consuming than the GA.} To the best of our knowledge, this is also the first work to investigate the impact of redundancy levels on fault detection capability under such constraints, providing valuable insights for future research and practical applications. Note that we retained FAST-R as one of the baselines, though it cannot simultaneously satisfy both the fixed-size minimization budget and the full coverage adequacy constraint. This is because FAST-R can be applied in two scenarios (i.e., fixed-size budget and adequate), which enables experimental comparison with our approach. 

\section{Conclusion}\label{sec:conclusion}

In this paper, we presented \ourapproach, a novel and practical approach for TSM guided by requirements coverage. As is often the case in domains where functional safety is crucial and must be demonstrated through adherence to standards, test cases are specified in natural language (test steps), traced to requirements, and implemented manually. 

\textcolor{black}{\ourapproach~combines the text embedding technique, distance function, and a GA to effectively reduce test suite size while strictly preserving full requirement coverage under a fixed budget constraint. We investigate three preprocessing methods for test cases, seven text embedding techniques, three distance functions, and three initialization strategies for the GA.}
Its evaluation on an industrial automotive dataset demonstrates its superior performance in \textit{FDR} across various minimization budgets. Compared with multiple baselines, \textcolor{black}{while being efficient in runtime, \ourapproach~achieves higher \textit{FDR} across most minimization budgets, making it a better choice in many practical contexts.}
Furthermore, our exploration of test suite redundancy levels revealed that higher redundancy significantly improves \textit{FDR}, while higher redundancy combined with low minimization budgets tends to significantly increase the risk of undetected faults, highlighting the importance of considering redundancy characteristics in TSM.  

In conclusion, our results establish \ourapproach~as a scalable and reliable minimization solution for requirement-based testing (validation) in safety-critical systems. In the future, we plan to apply \ourapproach~to a broader range of application scenarios to evaluate further and enhance its generalizability.

\begin{acks}
This work was supported by a research grant from Aptiv, the Discovery Grant and Canada Research Chair programs of the Natural Sciences and Engineering Research Council of Canada (NSERC), and a Research Ireland grant 13/RC/2094-2.
\end{acks}

\bibliographystyle{ACM-Reference-Format}


\end{document}